\documentclass[prl,twocolumn,showpacs,superscriptaddress,floatfix]{revtex4}
\usepackage{graphicx,amsfonts,amssymb,amsmath, hyperref}

\newif\ifhyper
\hypertrue
\ifhyper
\hypersetup{
   citecolor = {green},
   colorlinks = {true}, % false
   urlcolor = {blue} % magenta
}
\fi

\newcommand{\beq}{\begin{equation}}
\newcommand{\eeq}{\end{equation}}
\newcommand{\beqa}{\begin{eqnarray}}
\newcommand{\eeqa}{\end{eqnarray}}
\newcommand{\ket} [1] {\vert #1 \rangle}
\newcommand{\bra} [1] {\langle #1 \vert}
\newcommand{\braket}[2]{\langle #1 | #2 \rangle}

\begin{document}

\title{Simulation of two dimensional quantum systems on an infinite lattice revisited: \\
corner transfer matrix for tensor contraction}
\author{Rom\'an Or\'us}
\email{orus@physics.uq.edu.au}
\affiliation{School of Mathematics and Physics, The University of Queensland,
QLD 4072, Australia}

\author{Guifr\'e Vidal}
\affiliation{School of Mathematics and Physics, The University of Queensland,
QLD 4072, Australia}

\begin{abstract}
An extension of the \emph{projected entangled-pair states} (PEPS) algorithm to infinite systems, known as the iPEPS algorithm, was recently proposed to compute the ground state of quantum systems on an infinite two-dimensional lattice. Here we investigate a modification of the iPEPS algorithm, where the \emph{environment} is computed using the \emph{corner transfer matrix renormalization group} (CTMRG) method, instead of using \emph{one-dimensional transfer matrix} methods as in the original proposal. We describe a variant of the CTMRG that addresses different directions of the lattice independently, and use it combined with imaginary time evolution to compute the ground state of the two-dimensional quantum Ising model. Near criticality, the modified iPEPS algorithm is seen to provide a better estimation of the order parameter and correlators. 
%In the context of simulating quantum systems on an infinite two-dimensional lattice by means of \emph{infinite projected entangled pair states}, we investigate the performance of the \emph{corner transfer matrix renormalization group} (CTMRG) approach. Based on the corner transfer matrix (CTM) formalism, this approach is seen to be more robust than one-dimensional transfer matrix methods for the computation of local environments. We describe a variant of the CTMRG, the directional CTM, that addresses different directions of the lattice independently, and use it combined with imaginary time evolution to compute the ground state of the two-dimensional quantum Ising model. Accurate estimates of the critical magnetic field and $\beta$ exponent are obtained.
\end{abstract}


\maketitle

Understanding the emergent properties of many-body systems is one of the main goals of modern Physics. Projected entangled-pair states (PEPS) \cite{PEPS} were recently proposed by Verstraete and Cirac to describe the ground state of finite, inhomogeneous quantum systems on a 2D lattice. A PEPS consists of a two-dimensional network of tensors whose coefficients are optimized so as to approximate the ground state of a local Hamiltonian. An extension of the PEPS algorithm to infinite, homogeneous 2D lattices, known as infinite PEPS (iPEPS) algorithm \cite{iPEPS}, was also subsequently developed. As a many-body ansatz, the infinite PEPS had already been discussed by Sierra and Mart\'in-Delgado \cite{sierra} under the name of \emph{vertex matrix product ansatz}, and had been successfully used by Nishino and Okunishi \cite{TPS}, under the name of \emph{tensor product variational state}, to evaluate the partition function of a 3D classical system. In the context of 2D quantum systems, a simplified version of the ansatz, with only three free parameters, had also been used in Ref. \cite{tpva} prior to the iPEPS algorithm \cite{iPEPS}. By considering an evolution in imaginary time, the iPEPS algorithm overcame the stability problems of previous proposals while allowing for the optimization of all the coefficients in the ansatz. Subsequently, ingenious approaches to optimize homogeneous PEPS in finite systems with periodic boundary conditions have also been proposed, such as those in Refs. \cite{TERG} and \cite{Jiang}, where the ansatz is often called \emph{tensor product state}. So far, algorithms based on the PEPS formalism have provided accurate ground state properties of several models of spins and hard-core bosons \cite{tpva,PEPS,iPEPS,TERG,Jiang, Spins,compass,LiZhou,HardCore,Antiferro,Bela}. Importantly, they can address systems beyond the reach of quantum Monte Carlo, such as frustrated antiferromagnets \cite{Antiferro,Bela}.

An approximation $\ket{\Psi}$ to the ground state of a local Hamiltonian $H$ with an infinite PEPS is typically obtained either by minimizing the expected value of the energy $\bra{\Psi}H\ket{\Psi}$ or by simulating an evolution in imaginary time $\ket{\Psi} \approx e^{-H\tau}\ket{\Psi_0}$, where $\ket{\Psi_0}$ is some initial state. In either case, the tensors that define the infinite PEPS are optimized iteratively and, in order to properly update a given tensor, one needs to compute its \emph{environment}: a 2D tensor network that accounts for the rest of the tensors in the ansatz. Unfortunately, computing the environment is hard and an approximation scheme is required. In the case of a homogeneous system, several approximation schemes have been proposed:

(i) \emph{1D transfer matrix} approaches, such as the \emph{infinite time-evolving block decimation} (iTEBD) approach \cite{iTEBD} used in the original iPEPS algorithm \cite{iPEPS}. 

(ii) \emph{2D coarse-graining} approaches based on the tensor entanglement renormalization group (TERG) \cite{TERG, Jiang}.  

(iii) \emph{Corner transfer matrix} (CTM) approaches, such as the \emph{corner transfer matrix renormalization group} (CTMRG) algorithm \cite{CTMRG}. 

In this work we explore a modification of the iPEPS algorithm where instead of computing the environment using the iTEBD approach, as originally proposed in \cite{iPEPS}, we use the CTMRG \cite{CTMRG}. 

\begin{figure}
\includegraphics[width=0.45\textwidth]{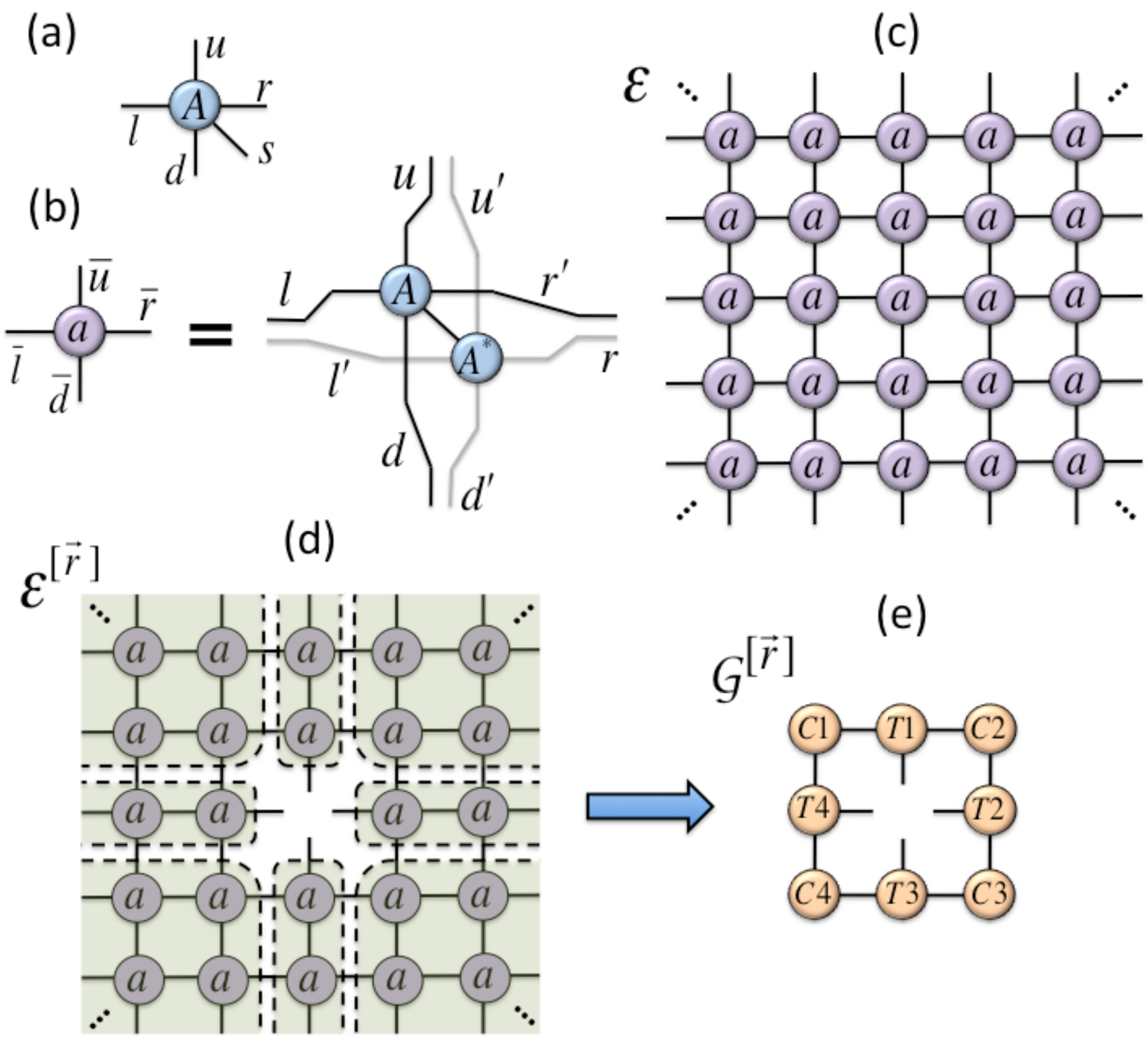}
 \caption{(color online) Diagrammatic representation of (a) infinite PEPS tensor $A$ with physical index $s$ and bond indices $u,r,d$ and $l$; (b) reduced tensor $a$; (c) infinite 2D tensor network $\mathcal{E}$; (d) environment $\mathcal{E}^{[\vec{r}]}$ for site $\vec{r}$; (e) eight-tensor effective environment $\mathcal{G}^{[\vec{r}]}$.} 
 \label{fig1}
 \end{figure} 

The CTM formalism was originally derived by Baxter \cite{CTM} and later adapted by Nishino and Okunishi in the CTMRG \cite{CTMRG} to numerically compute environments. The goal of this paper is to investigate the performance of CTM approaches to compute environments within the context of the iPEPS algorithm \cite{iPEPS}. Specifically, we first describe a versatile variant of the CTMRG, the directional CTM approach, which addresses different directions of the lattice separately, and use it in conjunction with imaginary time evolution to study the 2D quantum Ising model near criticality. Accurate estimates of the critical magnetic field and $\beta$ exponent are obtained. Then we compare the results obtained using the original iPEPS algorithm (where the environment was computed using the iTEBD \cite{iTEBD}) with this new version of the algorithm. The modified iPEPS algorithm is seen to converge signficantly faster to the ground state and provide a better characterization of the critical point \cite{classical}.
 
Let us consider an infinite PEPS for the state $\ket{\Psi}$ of an infinite 2D lattice $\mathcal{L}$, which for concreteness we take to be a square lattice, with each site labeled by two integers $\vec{r}=(x,y)$ and represented by a complex vector space $\mathbb{V}$ of dimension $d$. In the simplest scenario, the infinite PEPS is characterized by a single tensor $A$ that is repeated on all lattice sites. It has components $A_{s\ udlr}$, where $s$ labels a local (physical) basis of $\mathbb{V}$ ($s=1,\cdots,d$ ) and $u,d,l,r$ are bond indices ranging from $1$ to $D$, with $D$ the bond dimension of the infinite PEPS, see Fig. \ref{fig1}(a). Let $a$ denote the reduced tensor $a \equiv \sum_{s=1}^d A_s \otimes A_s^{*}$, with double bond indices such as $\bar{u} = (u,u')$ (see Fig. \ref{fig1}(b)). Then the scalar product $\braket{\Psi}{\Psi}$ can be expressed as a two-dimensional network $\mathcal{E}$ made of infinitely many copies of $a$, Fig. \ref{fig1}(c). [Notice that with a proper choice of tensor $a$, $\mathcal{E}$ can also represent the partition function of a 2D classical statistical model]. The environment of site $\vec{r}$, $\mathcal{E}^{[\vec{r}]}= \partial\mathcal{E}/\partial a^{[\vec{r}]}$, is obtained from $\mathcal{E}$ by removing the tensor $a$ on site $\vec{r}$, see Fig. \ref{fig1}(d). The goal of the CTMRG algorithm is to compute an approximation $\mathcal{G}^{[\vec{r}]}$ to $\mathcal{E}^{[\vec{r}]}$ by finding the fixed point of four CTMs. This effective environment is given in terms of a small tensor network, $\mathcal{G}^{[\vec{r}]} =\{C_1, T_1, C_2, T_2, C_3, T_3, C_4, T_4 \}$, where tensors $C_1, C_2, C_3, C_4$ represent four CTMs (one for each corner), and tensors  $T_1,T_2,T_3,T_4$ represent two half-column and two half-row transfer matrices, Fig. \ref{fig1}(e). 
 
\begin{figure}
 \includegraphics[width=0.42\textwidth]{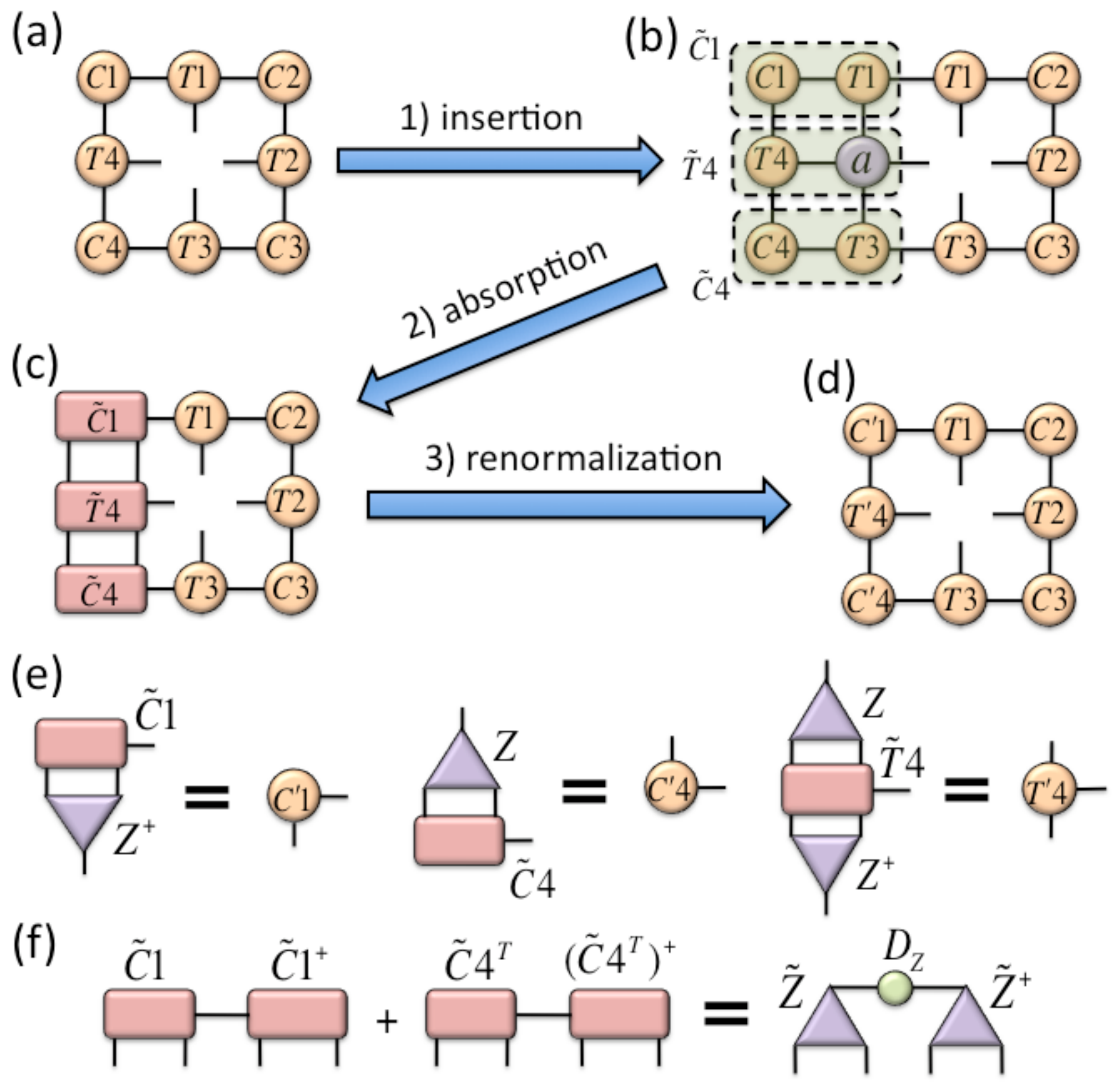}
 \caption{(color online) (a)-(d) Main steps of a left move: insertion, absorption and renormalization; (e) the CTMs $\widetilde{C}_1$, $\widetilde{C}_4$ and the half-row transfer matrix $\widetilde{T}_4$ are renormalized with isommetry $Z$; (f) eigenvalue decomposition for the sum of the squares of CTMs $\widetilde{C}_1$ and $\widetilde{C}_4$.} 
 \label{fig2}
 \end{figure}

In the directional variant of the CTMRG used in this work, the eight tensors of $\mathcal{G}^{[r]}$ are updated according to four \emph{directional coarse-graining moves}, namely left, right, up and down moves, which are iterated until the environment converges. Given an effective environment $\mathcal{G}^{[\vec{r}]} = \{C_1, T_1, C_2, T_2, C_3, T_3, C_4, T_4 \}$, a move, e.g. to the left, consists of the following three main steps, Fig. \ref{fig2}: 
\vspace{5pt}

\underline{(1) \emph{Insertion}}: insert a new column made of tensors $T_1$, $a$ and $T_3$ as in Fig. \ref{fig2}(b).

\underline{(2) \emph{Absorption}}: contract tensors $C_1$ and $T_1$, tensors $C_4$ and $T_3$, and also tensors $T_4$ and $a$, resulting in two new CTMs $\widetilde{C}_1$ and $\widetilde{C}_4$, and a new half-row transfer matrix $\widetilde{T}_4$, see Fig. \ref{fig2}(c). 

\underline{(3) \emph{Renormalization}}: Truncate the vertical indices of $\widetilde{C}_1$, $\widetilde{T}_4$ and $\widetilde{C}_4$ by inserting the isometry $Z$, $Z^{\dagger}Z = I$. This produces renormalized CTM's $C'_1 = Z^{\dagger} \widetilde{C}_1$, $C'_4 = \widetilde{C}_4 Z$ and half-row transfer matrix $T'_4$, Fig. \ref{fig2}(d)-(e).

\vspace{5pt}

A proper choice of isometry $Z$ in the renormalization step is of great importance. One possibility is to use the eigenvalue decomposition of the product of the four CTMs $\tilde{C}_1$, $C_2$, $C_3$, $\tilde{C}_4$ as in Ref. \cite{CTMRG}. Here we consider instead the eigenvalue decomposition of $\tilde{C}_1\tilde{C}_1^{\dagger} + \tilde{C}_4^{\dagger}\tilde{C}_4 = \tilde{Z} D_Z \tilde{Z}^{\dagger}$, Fig. \ref{fig2}(f), and use the isometry $Z$ that results from keeping the entries of $\tilde{Z}$ corresponding to the $\chi$ largest eigenvalues of $D_Z$. Similarly to Ref. \cite{CTMRG}, this isometry targets the CTMs of the effective environment instead of the wave function itself. The cost of implementing these steps scales with $D$ and $\chi$ as $O(D^6 \chi^3)$. The net result is a new effective environment $\mathcal{G}'^{[\vec{r}]}$ for site $\vec{r}$ given by tensors $\{C'_1, T_1, C_2, T_2, C_3, T_3, C'_4, T'_4 \}$, see Fig. \ref{fig2}(d). 
By composing the four moves of the directional CTM we recover one iteration of CTMRG \cite{CTMRG}. The additional flexibility provided by individual moves can be used to accelerate convergence in a specific direction, e.g. in highly anisotropic systems. In addition, the prescription used to compute the isometry $Z$ is still valid --and produces stable results \cite{classical}-- in the context of simulating imaginary time evolution described in this work. As with other similar methods \cite{iTEBD, CTMRG, Levin}, an immediate application of the directional CTM is to compute expected values from 2D classical partition functions (results not shown).
 
\begin{figure}
 \includegraphics[width=0.49\textwidth]{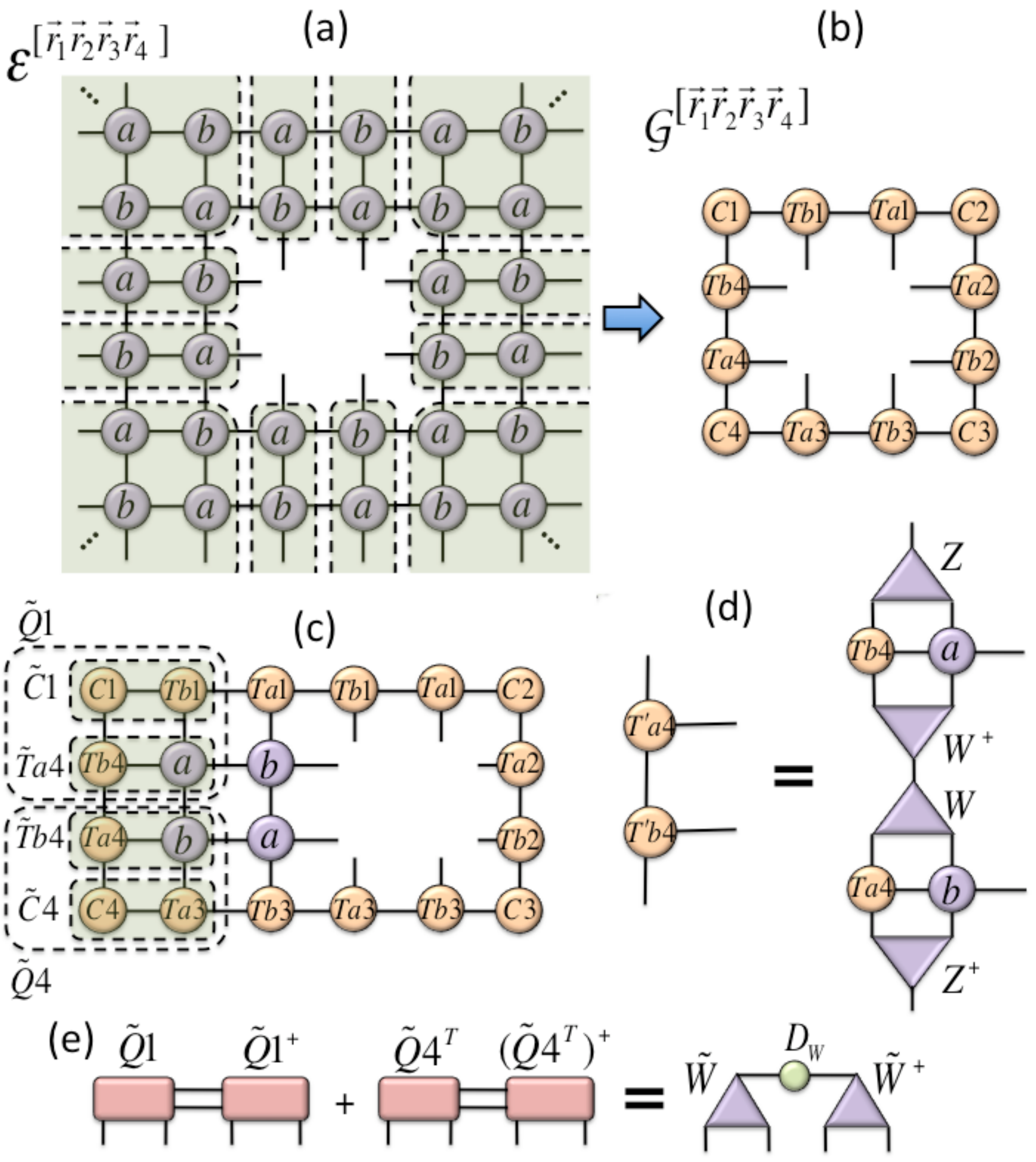}
 \caption{(color online) (a) Environment of the four-site unit cell; (b) twelve-tensor effective environment; (c) two new columns are inserted, and absorbed towards the left and renornalized individually. The diagram shows the contraction leading to $\widetilde{C}_1, \widetilde{C}_4$, $\widetilde{T}_{a4}$ and $\widetilde{T}_{b4}$  when absorbing the first column, and also to the CTMs $\widetilde{Q}_1$ and $\widetilde{Q}_4$; (d) two isometries $Z$ and $W$ are used to obtain the renormalized half-row transfer matrices $T'_{a4}$ and $T'_{b4}$; (e) eigenvalue decomposition for the sum of squares of CTMs $\widetilde{Q}_1$ and $\widetilde{Q}_4$.}
\label{fig4}
 \end{figure}

In order to compute the ground state of 2D quantum models by simulating imaginary time evolution, we consider an infinite PEPS characterized by two tensors $A$ and $B$. The optimization of tensors $A$ and $B$ proceeds in the same way as we proposed as part of the iPEPS algorithm \cite{iPEPS}, but with the crucial difference that here the required environment for two contiguous sites is computed with the directional CTM instead of using 1D transfer matrix (1DTM) techniques. The scalar product $\braket{\Psi}{\Psi}$ consists of an infinite 2D tensor network made of copies of the reduced tensors $a$ and $b$. We first consider the environment  $\mathcal{E}^{[\vec{r}_1, \vec{r}_2, \vec{r}_3, \vec{r}_4]}$ of a four-site unit cell, see Fig. \ref{fig4}(a), and approximate it with an effective environment $\mathcal{G}^{[\vec{r}_1, \vec{r}_2, \vec{r}_3, \vec{r}_4]}  = \{C_1, T_{b1}, T_{a1}, C_2, T_{a2}, T_{b2}, C_3, T_{b3}, T_{a3}, C_4, T_{a4}, T_{b4} \}$ made of twelve tensors, which are computed by iterating left, right, up and down moves. 
These directional moves are a natural adaptation to a four-tensor unit cell of those in Fig. \ref{fig2}. As shown in Fig. \ref{fig4}(b), the half-row and half-column transfer matrices $T_i$ ($i=1,2,3,4$) are replaced with pairs of half-row and half-colum transfer matrices $T_{ai}, T_{bi}$. This time, in order to implement e.g. a left move, \emph{two new columns} are inserted in the system in step (1), see Fig.\ref{fig4}(c). For each of the inserted columns, we perform the absorption and renormalization steps (2) and (3). The renormalization step requires introducing an additional isometry $W$, which we compute in an analogous way as isometry $Z$, see Fig. \ref{fig4}(e). As before, the cost of a move scales as $O(D^6\chi^3)$.
Finally, from a converged environment for the four-site unit cell, an effective environment for any pair of nearest neighbor sites is easily obtained with an additional directional move. 

To demonstrate the performance of the approach, we have computed an infinite PEPS approximation to the ground state of the spin-1/2 quantum Ising model on a transverse magnetic field, 
$H_I(\lambda) = -\sum_{\langle\vec{r},\vec{r}'\rangle} \sigma_z^{[\vec{r}]} \sigma_z^{[\vec{r}']} - \lambda \sum_{\vec{r}} \sigma_x^{[\vec{r}]}$, by simulating an imaginary time evolution. The simulation proceeds as in Ref. \cite{iPEPS}, but we use the directional CTM to obtain the effective environment at each step of the imaginary time evolution. This evolution is performed with decreasing time steps $\delta \tau$ ranging from $10^{-1}$ to $10^{-5}$, and until convergence of local observables and two-point correlators is attained.

\begin{figure}
 \includegraphics[width=0.49\textwidth]{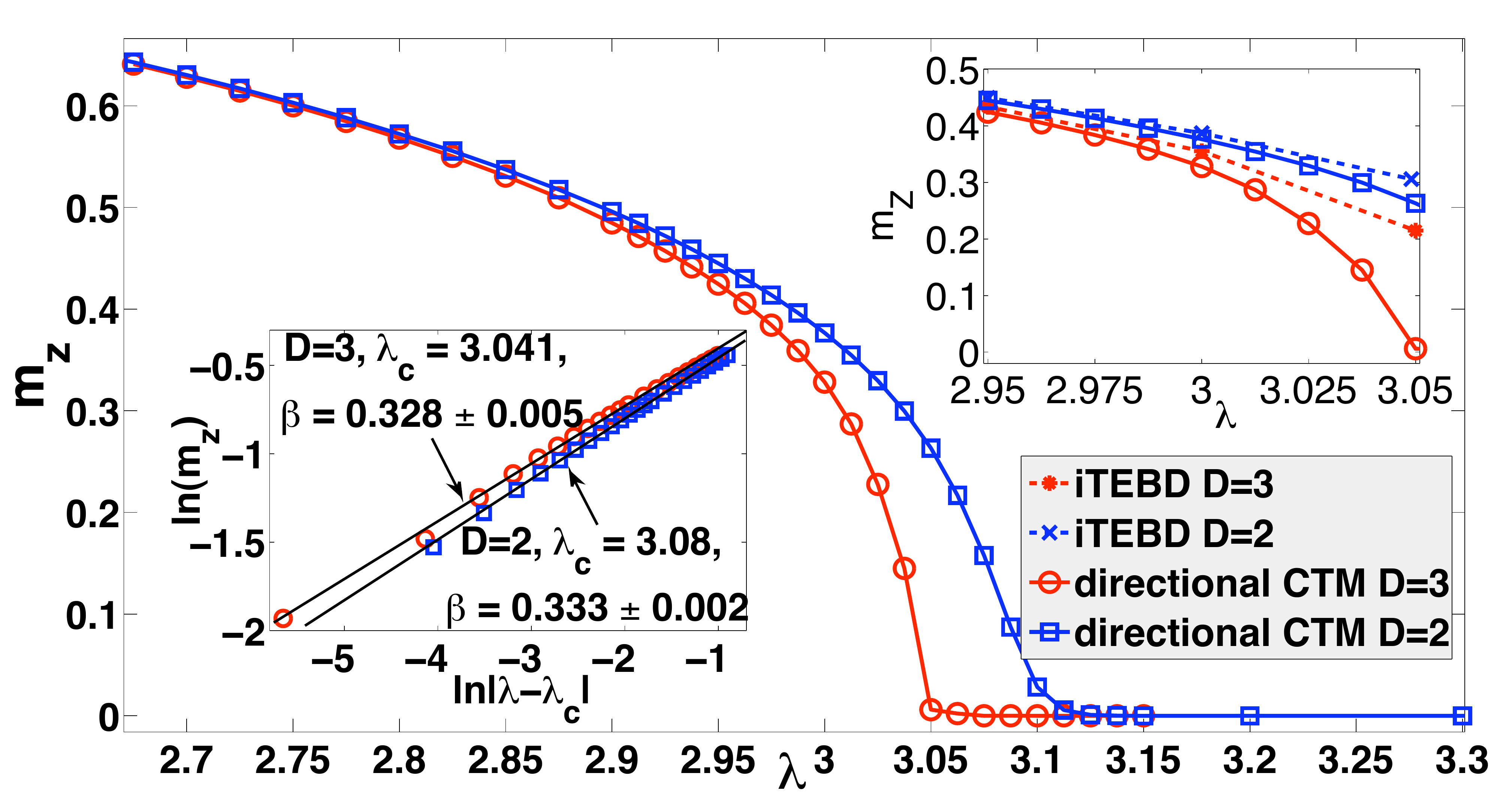}
 \caption{(color online) Order parameter $m_z$ as a function of the transverse field $\lambda$, computed with the directional CTM approach. Lines are a guide to the eye. The lower-left inset shows a log plot (in natural logarithms) of $m_z$ versus $| \lambda - \lambda_c|$, including our estimates for $\lambda_c$ and $\beta$. The continuous lines show the linear fits. The upper-right inset shows a comparison close to criticality with the results from Ref.{\cite{iPEPS}} using the original version of the iPEPS algorithm, which used iTEBD (dashed lines). Results correspond to $(D,\chi)$ equal to $(2,20)$ and $(3,30)$.}
 \label{fig5}
\end{figure}

Fig. \ref{fig5} shows the order parameter $m_z \equiv \bra{\Psi} \sigma_z \ket{\Psi}$ as a function of the transverse magnetic field $\lambda$, for $(D,\chi)$ equal to $(2,20)$ and $(3,30)$, where the value of $\chi$ is chosen so that the results are converged with respect to this parameter. Remarkably, an infinite PEPS with bond dimension $D=3$ already produces results within less than a percent from the best quantum Monte Carlo estimates for the critical magnetic field $\lambda_c^{MC} \approx 3.044$ and critical exponent for the order parameter $\beta^{MC} \approx 0.327$ \cite{mcIs}, namely with relative errors $\approx 0.1\%$ and $ \approx 0.3\%$ respectively. Table \ref{table1} contains a comparison with results obtained with the original version of the iPEPS algorithm and with the TERG algorithm for large systems \cite{TERG}. 

\begin{table}
	\begin{tabular}{|c| c |c |c| c| c|}
	\hline
	  & iPEPS with   & iPEPS with  & TERG {\scriptsize \cite{TERG}} \\
	  & directional CTM & iTEBD {\scriptsize \cite{iPEPS}} & \\
	\hline\hline
	$\lambda_c$ D=2 & 3.08 & 3.10 & 3.08  \\
	\;\;\; D=3 & 3.04 & 3.06& -  \\
	\hline
	$\beta$ D=2 &0.333 & 0.346 & 0.333   \\
	\;\;\; D=3 &0.328 & 0.332 & -   \\
	\hline 
	\end{tabular}
	\caption{Critical point $\lambda_c$ and exponent $\beta$ for the 2D quantum Ising model as estimated by the new and old versions of the iPEPS algorithm, as well as the TERG (for a finite lattice of up to $2^9 \times 2^9$ spins). For reference, the quantum Monte Carlo estimation is $\lambda^{MC}_c \approx 3.044$ and $\beta^{MC} \approx 0.327$ \cite{mcIs}.}
	\label{table1}
\end{table}

\begin{figure}
 \includegraphics[width=0.49\textwidth]{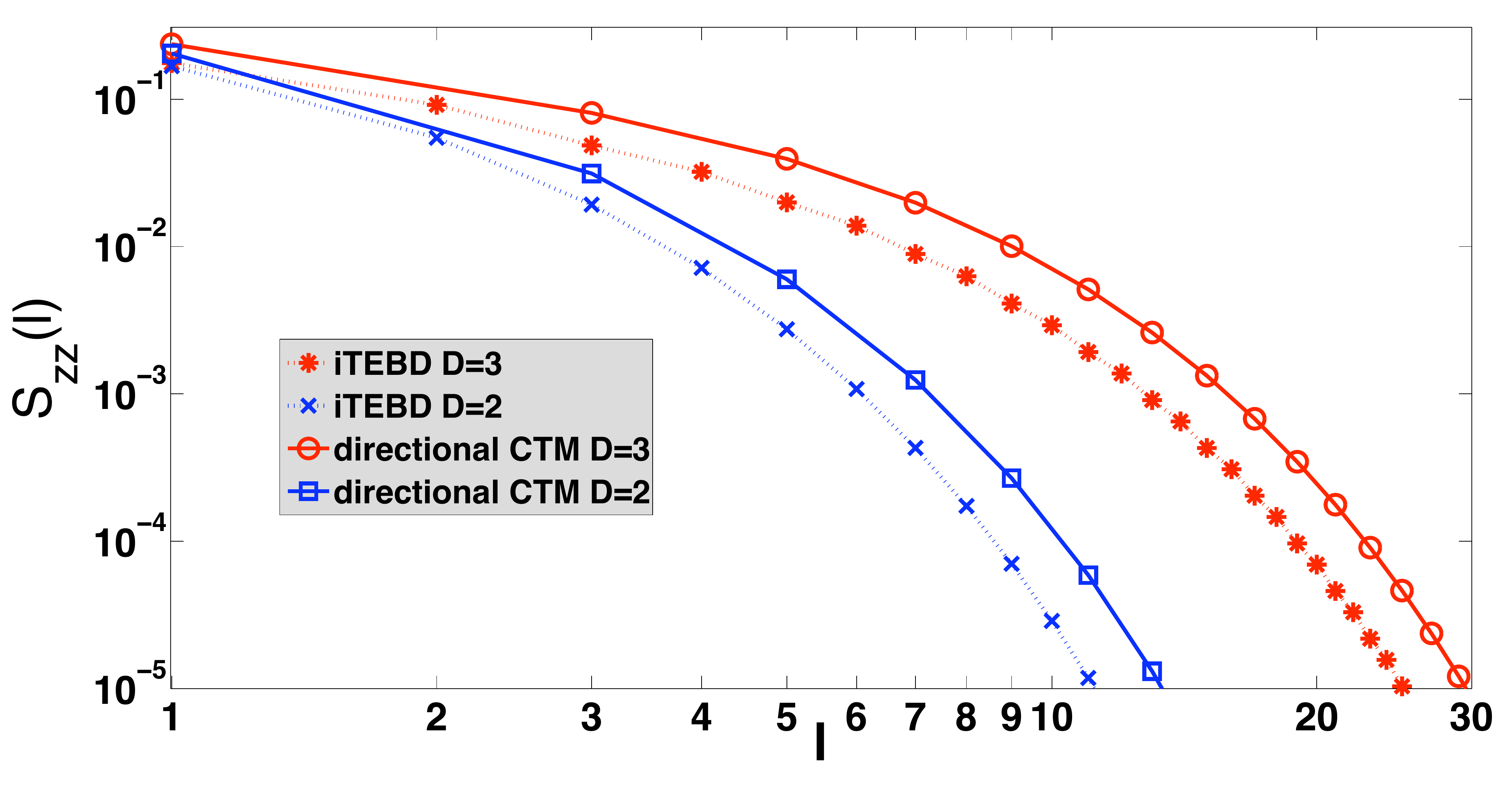}
 \caption{(color online) Log plot of the correlator $S_{zz}(l)$ (in base 10), as computed with the original and present versions of the iPEPS algorithm, namely using the iTEBD (dashed lines) and the directional CTM (solid lines) approaches. Lines are a guide to the eye. Our results are for the $(D,\chi)$ pairs $(2,20)$ (for $\lambda = 3.09$) and $(3,30)$ (for $\lambda = 3.04$).}
 \label{fig6}
\end{figure}

It is particularly instructive to compare the performance of the original and present versions of the iPEPS algorithm, since they are both based on imaginary time evolution and only differ in how the two-site environment is computed: by means of the iTEBD and directional CTM approaches, respectively. One finds that when computing environments using the directional CTM, a significantly better infinite PEPS approximation to near-critical ground states is obtained, leading to a more accurate characterization of the quantum phase transition. As shown in Fig. \ref{fig6}, the resulting infinite PEPS also displays stronger correlators $S_{zz}(l) \equiv \bra{\Psi} \sigma_z^{[\vec{r}]} \sigma_z^{[\vec{r} + l \hat{e}_x]} \ket{\Psi} - (m_z)^2$.

However, further comparison of results involving also other spin models reveals that, away from the quantum critical point, both the directional CTM and the iTEBD approaches yield equivalent accuracies for ground state properties. In particular, both versions of the iPEPS algorithm are equally suited to study first order phase transitions, a task for which they are particularly successful \cite{compass,Bela}. Indistinguishable results are also obtained in models with long or infinite correlation lengths, such as the Heisenberg antiferromagnet on a square lattice \cite{Bela}, when using a bond dimension $D$ that is too small to offer a proper approximation to the ground state (in Ref. \cite{Bela}, a bond dimension $D=5$ still produces a spontaneous magnetization that is off by $10\%$). It seems, therefore, that computing environments with a CTM approach leads to better results when the following two requirements are simultaneously met: (i) the ground state must have a long correlation length (e.g. near a quantum critical point), and (ii) the bond dimension $D$ must be sufficiently large that the ansatz can in principle properly approximate the ground state.

The correlators in Fig. \ref{fig6} also show that numerical calculations with infinite PEPS introduce an artificial finite correlation length at criticality. This is a consequence of the truncation in parameter $\chi$ in the calculations of the effective environments. This truncation effect is similar to the one discussed in Ref. \cite{luca} and could in principle be analyzed in a systematic way. 
%However, such an analysis is beyond the scope of the present work. 

To summarize, PEPS are a valuable ansatz to approximate the ground state of 2D lattice models, as previously demonstrated by several authors \cite{tpva,PEPS,iPEPS,TERG,Jiang, Spins,compass,LiZhou,HardCore,Antiferro,Bela} with a variety of systems of interacting spins and hard-core bosons, including frustrated spins \cite{Antiferro,Bela} that cannot be addressed with quantum Monte Carlo techniques. Several methods have been proposed to optimize the PEPS. The main factor limiting the accuracy of the results is the bond dimension $D$. However, for a fixed value of the bond dimension $D$, the quality of the approximation may also depend on the method used to optimize the ansatz. In this work we have investigated a modification of the iPEPS algorithm, where the iTEBD of the original proposal has been replaced with a directional CTM, a variant of CTMRG \cite{CTMRG}. The new version of the algorithm provides a significantly better description of the ground state near criticality.

{\it Acknowledgements.--} Suppoert from the University of Queensland (ECR2007002059) and the Australian Research Council (FF0668731, DP0878830) is acknowledged.

{}


\begin{thebibliography}{}
\bibitem{classical} The accuracies obtained in this work for the critical point of the 2D quantum Ising model are comparable to those obtained in Ref. \cite{TPVA} for the 3D classical Ising model. The present approach, also valid for 2D quantum systems without 3D classical counterpart, does not suffer from the instabilities reported in Ref. \cite{TPVA}.
\end{thebibliography}
\end{document}